\documentclass[twocolumn,english,pre,aps,superscriptaddress,showpacs]{revtex4}
\usepackage[T1]{fontenc}
\usepackage[latin1]{inputenc}
\usepackage{graphicx}

\makeatletter

\def\vec#1{\mbox{\boldmath $\mathrm{#1}$}}
\def\Pm{\textrm{Pm}}
\def\Rm{\textrm{Rm}}
\def\RE{\textrm{Re}}
\def\Ha{\textrm{Ha}}

\usepackage{babel}
\makeatother
\begin{document}

\title{Paradox of inductionless magnetorotational instability in a Taylor-Couette
flow with a helical magnetic field}

\author{J\={a}nis Priede }

\affiliation{Institute of Physics, University of Latvia, LV--2169 Salaspils, Latvia}

\email{priede@sal.lv}

\affiliation{Forschungszentrum Rossendorf, MHD Department, P.O. Box 510119, D--01314
Dresden, Germany}

\author{Ilm\={a}rs Grants}

\author{Gunter Gerbeth}

\affiliation{Forschungszentrum Rossendorf, MHD Department, P.O. Box 510119, D--01314
Dresden, Germany}

\begin{abstract}
We consider the magnetorotational instability (MRI) of a
hydrodynamically stable Taylor-Couette flow with a helical
external magnetic field in the inductionless approximation defined
by a zero magnetic Prandtl number ($\Pm=0)$. This leads to a
considerable simplification of the problem eventually containing
only hydrodynamic variables. First, we point out that the energy
of any perturbation growing in the presence of magnetic field has
to grow faster without the field. This is a paradox because the
base flow is stable without the magnetic while it is unstable in
the presence of a helical magnetic field without being modified by
the latter as it has been found recently by Hollerbach and
R\"{u}diger {[}Phys. Rev. Lett. \textbf{95}, 124501 (2005){]}. We
revisit this problem by using a Chebyshev collocation method to
calculate the eigenvalue spectrum of the linearized problem. In
this way, we confirm that MRI with helical magnetic field indeed
works in the inductionless limit where the destabilization effect
appears as an effective shift of the Rayleigh line. Second, we
integrate the linearized equations in time to study the transient
behavior of small amplitude perturbations, thus showing that the
energy arguments are correct as well. However, there is no real
contradiction between both facts. The linear stability theory
predicts the asymptotic development of an arbitrary
small-amplitude perturbation, while the energy stability theory
yields the instant growth rate of any particular perturbation, but
it does not account for the evolution of this perturbation. Thus,
although switching off the magnetic field instantly increases the
energy growth rate, in the same time the critical perturbation
ceases to be an eigenmode without the magnetic field.
Consequently, this perturbation is transformed with time and so
looses its ability to extract energy from the base flow necessary
for the growth.
\end{abstract}


\maketitle The magnetorotational instability (MRI) is thought to
be responsible for the fast formation of stars and entire galaxies
in accretion disks. For a star to form, the matter rotating around
it has to slow down by transferring its angular momentum outwards.
Without MRI this process would take much longer than observed
because the velocity distribution in the accretion disks seems to
be hydrodynamically stable while the viscosity alone is not
sufficient to account for the actual accretion rates. It was
suggested by Balbus and Hawley \cite{Balbus-Hawley-1991} that a
Keplerian velocity distribution in an accretion disk can be
destabilized by a magnetic field analogously to a hydrodynamically
stable cylindrical Taylor-Couette flow as it was originally found
by Velikhov \cite{Velikhov-1959} and later analysed in more detail
by Chandrasekhar \cite{Chandrasekhar-1960}. In this case, the
effect of {}``frowziness'' of the axial magnetic field in a well
conducting fluid provides an additional coupling between the
meridional and azimuthal flow perturbations that, however,
requires a magnetic Reynolds number of $\Rm\sim10.$ For a liquid
metal with the magnetic Prandtl number $\Pm\sim10^{-5}-10^{-6}$
this corresponds to a hydrodynamic Reynolds number
$\RE=\Rm/\Pm\sim10^{6}-10^{7}$
\cite{Goodman-Ji-2002,Ruediger-etal-2003}. Thus, this instability
is hardly observable in the laboratory because any conceivable
flow at such Reynolds number would be turbulent. However, it was
shown recently by Hollerbach and R\"{u}diger
\cite{Hollerbach-Ruediger-2005} that MRI can take place in the
Taylor-Couette flow at $\RE\sim10^{3}$ when the imposed magnetic
field is helical. The most surprising fact is that this type of
MRI persists even in the inductionless limit of $\Pm=0$ where the
critical Reynolds number of the conventional MRI diverges as
$\sim1/\Pm.$ This limit of $\Pm=0$ formally corresponds to a
poorly conducting medium where the induced currents are so weak
that their magnetic field is negligible with respect to the
imposed field. Thus, on one hand, the imposed magnetic field does
not affect the base flow, which is the only source of energy for
the perturbation growth. But on the other hand, perturbations are
subject to additional damping due to the Ohmic dissipation caused
by the induced currents.

We show rigorously that the imposed magnetic field indeed reduces
the energy growth rate of any particular perturbation. On one hand,
this implies that the energy of any perturbation, which is growing
in the presence of magnetic field, has to grow even faster without
the field and vice versa. But on the other hand, the flow which is
found to be unstable in the presence of magnetic field is certainly
known to be stable without the field. This apparent contradiction
constitutes the paradox of the inductionless MRI which we address
in this study.

Consider an incompressible fluid of kinematic viscosity $\nu$ and
electrical conductivity $\sigma$ filling the gap between two infinite
concentric cylinders with inner radius $R_{i}$ and outer radius $R_{o}$
rotating with angular velocities $\Omega_{i}$ and $\Omega_{o}$,
respectively, in the presence of an externally imposed steady magnetic
field $\vec{B}_{0}=B_{\phi}\vec{e}_{\phi}+B_{z}\vec{e}_{z}$ with
axial and azimuthal components $B_{z}=B_{0}$ and $B_{\phi}=\beta B_{0}R_{i}/r$
in cylindrical coordinates $(r,\phi,z),$ where $\beta$ is a dimensionless
parameter characterizing the geometrical helicity of the field. Further,
we assume the magnetic field of the currents induced by the fluid
flow to be negligible relative to the imposed field that corresponds
to the so-called inductionless approximation holding for most of liquid-metal
magnetohydrodynamics characterized by small magnetic Reynolds numbers
$\textrm{Rm}=\mu_{0}\sigma v_{0}L\ll1,$ where $\mu_{0}$ is the magnetic
permeability of vacuum, $v_{0}$ and $L$ are the characteristic velocity
and length scale. The velocity of fluid flow $\vec{v}$ is governed
by the Navier-Stokes equation with electromagnetic body force \begin{equation}
\frac{\partial\vec{v}}{\partial t}+(\vec{v}\cdot\vec{\nabla})\vec{v}=-\frac{1}{\rho}\vec{\nabla}p+\nu\vec{\nabla}^{2}\vec{v}+\frac{1}{\rho}\vec{j}\times\vec{B}_{0},\label{eq:N-S}\end{equation}
 where the induced current follows from Ohm's law for a moving medium
\begin{equation}
\vec{j}=\sigma\left(\vec{E}+\vec{v}\times\vec{B}_{0}\right).\label{eq:Ohm}\end{equation}
 In addition, we assume that the characteristic time of velocity variation
is much longer than the magnetic diffusion time $\tau_{0}\gg\tau_{m}=\mu_{0}\sigma L^{2}$
that leads to the quasi-stationary approximation, according to which
$\vec{\nabla}\times\vec{E}=0$ and $\vec{E}=-\vec{\nabla}\Phi,$ where
$\Phi$ is the electrostatic potential. Mass and charge conservation
imply $\vec{\nabla}\cdot\vec{v}=\vec{\nabla}\cdot\vec{j}=0.$

The problem admits a base state with a purely azimuthal velocity distribution
$\vec{v}_{0}(r)=\vec{e}_{\phi}v_{0}(r),$ where \[
v_{0}(r)=r\frac{\Omega_{o}R_{o}^{2}-\Omega_{i}R_{i}^{2}}{R_{o}^{2}-R_{i}^{2}}+\frac{1}{r}\frac{\Omega_{o}-\Omega_{i}}{R_{o}^{-2}-R_{i}^{-2}}.\]
 Note that the magnetic field does not affect the base flow because
it gives rise only to the electrostatic potential $\Phi_{0}(r)=B_{0}\int v_{0}(r)dr$
whose gradient compensates the induced electric field so that there
is no current in the base state $(\vec{j}_{0}=0)$. However, a current
may appear in a perturbed state \[
\left\{ \begin{array}{c}
\vec{v},p\\
\vec{j},\Phi\end{array}\right\} (\vec{r},t)=\left\{ \begin{array}{c}
\vec{v}_{0},p_{0}\\
\vec{j}_{0},\Phi_{0}\end{array}\right\} (r)+\left\{ \begin{array}{c}
\vec{v}_{1},p_{1}\\
\vec{j}_{1},\Phi_{1}\end{array}\right\} (\vec{r},t)\]
 where $\vec{v}_{1},$ $p_{1},$ $\vec{j}_{1},$ and $\Phi_{1}$ present
small-amplitude perturbations for which Eqs. (\ref{eq:N-S}, \ref{eq:Ohm})
after linearization take the form \begin{eqnarray}
\frac{\partial\vec{v}_{1}}{\partial t} & + & (\vec{v}_{1}\cdot\vec{\nabla})\vec{v}_{0}+(\vec{v}_{0}\cdot\vec{\nabla})\vec{v}_{1}\nonumber \\
 & = & -\frac{1}{\rho}\vec{\nabla}p_{1}+\nu\vec{\nabla}^{2}\vec{v}_{1}+\frac{1}{\rho}\vec{j}_{1}\times\vec{B}_{0}\label{eq:v1}\\
\vec{j}_{1} & = & \sigma\left(-\vec{\nabla}\Phi_{1}+\vec{v}_{1}\times\vec{B}_{0}\right).\label{eq:j1}\end{eqnarray}
 In the following, we focus on axisymmetric perturbations, which are
typically much more unstable than non-axisymmetric ones \cite{Rued-ANN}.
In this case, the solenoidity constraints are satisfied by meridional
stream functions for fluid flow and electric current as \[
\vec{v}=v\vec{e}_{\phi}+\vec{\nabla}\times(\psi\vec{e}_{\phi}),\qquad\vec{j}=j\vec{e}_{\phi}+\vec{\nabla}\times(h\vec{e}_{\phi}).\]
 Note that $h$ is the azimuthal component of the induced magnetic
field which is used subsequently as an alternative to $\Phi$ for
the description of the induced current. In addition, for numerical
purposes, we introduce also the vorticity $\vec{\omega}=\omega\vec{e}_{\phi}+\vec{\nabla}\times(v\vec{e}_{\phi})=\vec{\nabla}\times\vec{v}$
as an auxiliary variable. Then the perturbation may be sought in the
normal mode form \[
\left\{ v_{1},\omega_{1,}\psi_{1},h_{1}\right\} (\vec{r},t)=\left\{ \hat{v},\hat{\omega},\hat{\psi},\hat{h}\right\} (r)\times e^{\gamma t+ikz},\]
 where $\gamma$ is in general a complex growth rate and $k$ is the
axial wave number. Henceforth, we proceed to dimensionless variables
by using $R_{i},$ $R_{i}^{2}/\nu,$ $R_{i}\Omega_{i},$ $B_{0},$
and $\sigma B_{0}R_{i}\Omega_{i}$ as the length, time, velocity,
magnetic field, and current scales, respectively. The nondimensionalized
governing equations read as \begin{eqnarray}
\gamma\hat{v} & = & D_{k}\hat{v}+\RE ik\left(r^{2}\Omega\right)'r^{-1}\hat{\psi}+\Ha^{2}ik\hat{h},\label{eq:vhat}\\
\gamma\hat{\omega} & = & D_{k}\hat{\omega}+2\RE ik\Omega\hat{v}-\Ha^{2}ik\left(ik\hat{\psi}+2\beta r^{-2}\hat{h}\right),\label{eq:omghat}\\
0 & = & D_{k}\hat{\psi}+\hat{\omega},\label{eq:psihat}\\
0 & = & D_{k}\hat{h}+ik\left(\hat{v}-2\beta r^{-2}\hat{\psi}\right),\label{eq:hhat}\end{eqnarray}
 where $D_{k}f\equiv r^{-1}\left(rf'\right)'-(r^{-2}+k^{2})f$ and
the prime stands for $d/dr;$ $\RE=R_{i}^{2}\Omega_{i}/\nu$ and $\Ha=R_{i}B_{0}\sqrt{\sigma/(\rho\nu)}$
are Reynolds and Hartmann numbers, respectively; \[
\Omega(r)=\frac{\lambda^{-2}-\mu+r^{-2}\left(\mu-1\right)}{\lambda^{-2}-1}\]
 is the dimensionless angular velocity of the base flow defined using
$\lambda=R_{o}/R_{i}$ and $\mu=\Omega_{o}/\Omega_{i}$. Boundary
conditions for the flow perturbation on the inner and outer cylinders
at $r=1$ and $r=\lambda,$ respectively, are $\hat{v}=\hat{\psi}=\hat{\psi}'=0.$
Boundary conditions for $\hat{h}$ on insulating and perfectly conducting
cylinders, respectively, are $\hat{h}=0$ and $(r\hat{h})'=0$ at
$r=1;\lambda.$

The governing Eqs. (\ref{eq:vhat}--\ref{eq:hhat}) for perturbation
amplitudes were discretized using a spectral collocation method on
a Chebyshev-Lobatto grid with a typical number of internal points
$N=32-96$. Auxiliary Dirichlet boundary conditions for $\hat{\omega}$
were introduced and then numerically eliminated to satisfy the no-slip
boundary conditions $\hat{\psi}'=0.$ Electric stream function $\hat{h}$
was expressed in terms of $\hat{v}$ and $\hat{\psi}$ by solving
Eq. (\ref{eq:hhat}) and then substituted in Eqs. (\ref{eq:vhat},
\ref{eq:omghat}) that eventually resulted in the $2N\times2N$ complex
matrix eigenvalue problem which was efficiently solved by the LAPACK's
ZGEEV routine.

In addition, Eqs. (\ref{eq:vhat}--\ref{eq:hhat}) were discretized
by using a Chebyshev-tau approximation and integrated forward in
time by a fully implicit 2nd order scheme with linear
extrapolation of convective and magnetic terms. We tested the
numerical code by finding a few leading eigenmodes and eigenvalues
by the so-called {}``snapshot'' method \cite{Goldhirsch-etal} and
compared to the results of the above described code as well as to
the linear instability results \cite{Chandrasekhar-book} and
\cite{Hollerbach-Ruediger-2005}. Agreement was at least three
significant digits.

Equations (\ref{eq:v1}, \ref{eq:j1}) straightforwardly lead to the
kinetic energy variation rate of a virtual perturbation
$\vec{v}_{1}$ satisfying the incompressiblity constraint and the
boundary conditions. Multiplying Eq. (\ref{eq:v1}) scalarly by
$\vec{v}_{1}$ and then integrating over the volume $V$ which
extends axially over the perturbation wavelength, we obtain
\[
\frac{\partial E_{1}}{\partial
t}=-\int\left[\left(\vec{v}_{1}\cdot\vec{\nabla}\right)\vec{v}_{1}\right]\cdot\vec{v}_{0}dV-\int\left
(\nu\vec{\omega}_{1}^{2}+\frac{\vec{j}_{1}^{2}}{\sigma}\right)dV,
\]
 where $E_{1}=\frac{1}{2}\int\vec{v}_{1}^{2}dV$ is the energy of
perturbation. The first integral in the equation above accounts
for the interaction of the perturbation with the basic flow which
is not affected by the magnetic field as noted above. The sign of
this integral may vary depending on $\vec{v}_{1}.$ Thus, this term
presents a potential source of energy. In contrast, the second
term is negative definite presenting an energy sink due to both
viscosity and conductivity. Since the current is induced only in
the presence of a magnetic field while the source term does not
depend on the magnetic field, we conclude that the instant growth
rate of any given perturbation has to be lower with magnetic field
than without it \begin{equation} \left.\frac{\partial
E_{1}}{\partial t}\right|_{B_{0}>0}<\left.\frac{\partial
E_{1}}{\partial t}\right|_{B_{0}=0}.\label{eq:nrg}\end{equation}

\begin{figure}
\begin{center}\includegraphics[%
  width=0.49\textwidth]{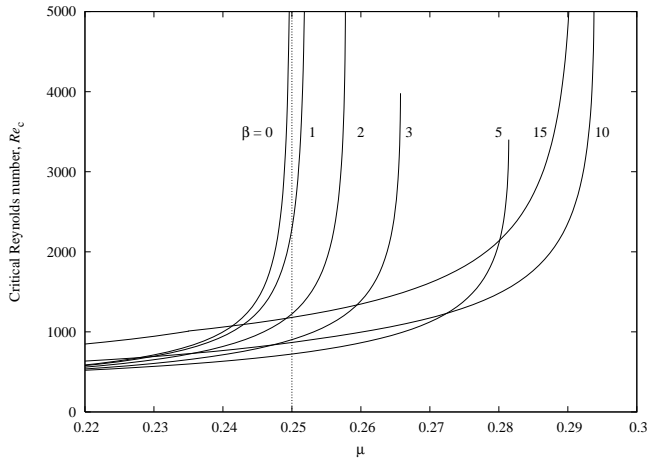}\end{center}

\caption{\label{cap:Rec-mu}Critical Reynolds number versus $\mu$ for insulating
cylinders with $\lambda=2$ at various helicities $\beta$ and fixed
Hartmann number $\Ha=15.$}
\end{figure}

\begin{figure}
\begin{center}\includegraphics[%
  width=0.5\textwidth]{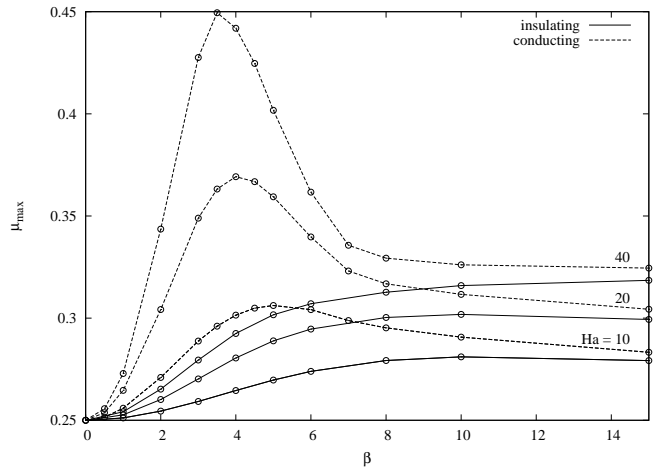}\end{center}

\caption{\label{cap:mu-beta}Limiting value of $\mu$ versus the helicity
$\beta$ for insulating and perfectly conducting cylinders with $\lambda=2$
at various Hartmann numbers.}
\end{figure}

The following results concern cylinders with $\lambda=2$, as in
Ref. \cite{Hollerbach-Ruediger-2005}. As seen in Fig.
\ref{cap:Rec-mu}, which shows the critical Reynolds number as a
function of $\mu$ for Hartmann number $\Ha=15$ and various
geometrical helicities $\beta,$ the linear instability threshold
can indeed extend well beyond the Rayleigh line
$\mu_{c}=\lambda^{-2}=0.25,$ defined by
$d\left(r^{2}\Omega\right)/dr=0$, when the magnetic field is
helical $(\beta\not=0).$ In contrast to $\Pm\not=0$
\cite{Hollerbach-Ruediger-2005}, the range of instability is
limited by $\mu_{\max},$ which is plotted in Fig.
\ref{cap:mu-beta} depending on the geometrical helicity $\beta$ at
various Hartmann numbers $\Ha$ for both insulating and perfectly
conducing cylinders. The critical $\RE$ tends to infinity as $\mu$
approaches $\mu_{\max}$ as in the nonmagnetic Taylor-Couette
instability. Thus, in the inductionless approximation, the
destabilizing effect of a helical magnetic field appears as a
shift of the Rayleigh line towards higher $\mu.$ The shift is
especially pronounced for perfectly conducting cylinders at
$\beta\approx4.$

\begin{figure}
\begin{center}\includegraphics[%
  width=0.5\textwidth]{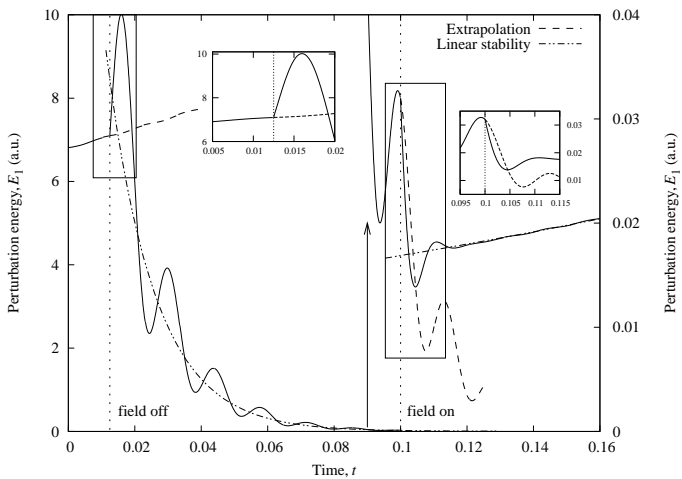}\end{center}

\caption{\label{cap:nrg}Time evolution of the energy of the
dominating perturbation with $k=2$ at $\RE=2000$ after switching
the magnetic field off and later on again for $\mu=0.27$,
$\Ha=15$, and $\beta=4$. Extrapolation shows how the evolution
would proceed without the change of the magnetic field. }
\end{figure}

The results of time-integration of the linearized problem are
illustrated in Fig. \ref{cap:nrg} for a perturbation with $k=2$ at
$\RE=2000.$ This perturbation is unstable in the presence of a
magnetic field with $\beta=4$ and $\Ha=15$ $(\RE_{c}=1554,$
$k_{c}=2.5)$ and stable without the field because
$\mu=0.27>\mu_{c}.$ First, we integrate an arbitrary, sufficiently
small initial perturbation for a sufficiently long time so that
the unstable mode dominates but still remains small for the linear
approximation to be valid. Then we {}``switch off'' the magnetic
field by setting $\Ha=0.$ Note that we assume the field to be
instantly absent when it is switched off. So we just compare the
evolution of the given perturbation with and without the field. As
seen on the first inset of Fig. \ref{cap:nrg}, the energy of an
unstable perturbation indeed starts to grow faster instantly after
the magnetic field is switched off. However, the growth keeps only
for a short time and then the energy quickly decays as predicted
by the linear stability analysis. Note that the energy keeps
decaying in an oscillatory way because the dominating perturbation
without the field is not a pure traveling wave but rather a
superposition of two oppositely traveling waves which both have
the same decay rate and frequency whereas the amplitude ratio of
both waves is determined by the initial condition.

The magnetic field is switched on again at the instant $t=0.1.$ The
corresponding evolution of the perturbation energy is shown on the
r.h.s. of Fig. \ref{cap:nrg} in enlarged scale. As seen in the second
inset, the energy decay rate instantly increases in accordance to
(\ref{eq:nrg}) when the magnetic field is switched on. However, after
a short transient the perturbation energy resumes the growth in agreement
with the linear stability analysis. In this case, the energy growth
is purely exponential because the dominating perturbation is a single
traveling wave.

Thus, this particular example of time integration confirms the
validity of Eq. (\ref{eq:nrg}) which applies in general to any
arbitrary perturbation. The energy of an unstable perturbation
indeed starts to grow faster when the magnetic field is switched
off. However, there is no real contradiction with the linear
stability predictions because the energy grows only for a limited
time and then turns to decay as predicted by the linear stability.
It is important to stress that the linear stability theory
predicts the asymptotic development of an arbitrary
small-amplitude perturbation, while the energy stability theory
yields the instant growth rate of any particular perturbation, but
it does not account for the evolution of this perturbation. Thus,
although switching off the magnetic field instantly increases the
energy growth rate of the most unstable as well as that of any
other perturbation, in the same time the critical perturbation
ceases to be an eigenmode without the magnetic field.
Consequently, this perturbation is transformed with time and so
looses its ability to extract energy from the base flow necessary
for the growth. Analogously, switching on the magnetic field
causes an instant decrease of the growth rate of any particular
perturbation because of Ohmic dissipation, while the magnetic
field transforms the perturbation so that it becomes able to
extract more energy from the base flow and so eventually grows.

To understand the physical mechanism of this instability, note
that a helical magnetic field, in contrast to pure axial or
azimuthal fields, provides an additional coupling between
meridional and azimuthal flow perturbations. In a helical magnetic
field with axial and azimuthal components, the radial component of
the meridional flow perturbation induces azimuthal and axial
current components, respectively. Interaction of this current with
the imposed magnetic field results in a purely radial
electromagnetic force which retards the original perturbation. So,
it has a stabilizing effect similar to the radial deformation of
magnetic flux lines in the conventional MRI
\cite{Velikhov-1959,Chandrasekhar-1960}. However, in the
perturbation of finite wavelength there is also a radial current
component associated with the axial one as required by the
solenoidity constraint. This radial current interacting with the
axial component of the helical magnetic field gives rise to the
azimuthal electromagnetic force, thus coupling the meridional and
azimuthal flow perturbations similarly to the conservation of the
angular momentum in the purely hydrodynamic Taylor-Couette
instability or the azimuthal twisting of axial magnetic flux lines
in the conventional MRI. Note that the latter effect also renders
the imposed axial magnetic field locally helical that, however,
requires $\Pm>0$ and $\RE\sim1/\Pm.$ When the imposed magnetic
field is helical, the inductionless approximation defined by
$\Pm=0$ is applicable to MRI where it leads to a considerable
simplification of the problem containing only hydrodynamic
variables as in the classical Taylor-Couette problem.

The research was supported by Deutsche Forschungsgemeinschaft in
frame of the Collaborative Research Centre SFB 609. The authors
would like to thank G. R\"udiger for helpful comments and
discussions.

\end{document}